\documentclass{aa}  

\usepackage{txfonts}
\usepackage[dvipsnames]{xcolor}
\usepackage{graphicx}
\usepackage{multirow}
\usepackage[shortlabels]{enumitem}
\usepackage{ulem}
\usepackage{soul}
\usepackage{txfonts}
\usepackage{subfigure}
\usepackage[
  breaklinks = true,
  colorlinks = true,
  urlcolor   = blue,
  citecolor  = blue,
  linkcolor  = blue,
]{hyperref}
\def\Halpha{\mbox{H\hspace{0.1ex}$\alpha$}}

\begin{document}

\title{Active chromospheric fibril singularity}

\subtitle{Coordinated observations from Solar Orbiter, SST, and IRIS}

    \author{Reetika Joshi\inst{1,2,3,4}
    \and
    Luc Rouppe van der Voort\inst{1,2}
    \and
    Guillaume Aulanier\inst{5,1}
    \and
    Sanja Danilovic\inst{6}
    \and
    Avijeet Prasad\inst{7,1,2}
    \and \\
    Carlos  J. Díaz Baso\inst{1,2}
    \and
    Daniel Nóbrega-Siverio\inst{8,9,1,2}
    \and
    Nicolas Poirier \inst{1,2}
    \and
    Daniele Calchetti \inst{10}
    }
          
    \institute
    {Rosseland Centre for Solar Physics, University of Oslo, P.O. Box 1029 Blindern, N-0315 Oslo, Norway\\
    \email{reetika.joshi@nasa.gov}  
    \and
    Institute of Theoretical Astrophysics, University of Oslo, P.O. Box 1029 Blindern, N-0315 Oslo, Norway
    \and
    NASA Goddard Space Flight Center, Heliophysics Science Division, Greenbelt, MD 20771, USA  
    \and
    Department of Physics and Astronomy, George Mason University, Fairfax, VA 22030, USA 
    \and
    Sorbonne Universit\'e, Observatoire de Paris - PSL, \'Ecole Polytechnique, Institut Polytechnique de Paris, CNRS, Laboratoire de physique des plasmas (LPP), 4 place Jussieu, F-75005 Paris, France
    \and 
    Institute for Solar Physics, Dept. of Astronomy, Stockholm University, Albanova University Center, 10691 Stockholm, Sweden
    \and
    Statkraft AS, Lysaker, Norway
    \and
    Instituto de Astrof\'isica de Canarias, E-38205 La Laguna, Tenerife, Spain
    \and
    Universidad de La Laguna, Dept. Astrof\'isica, E-38206 La  Laguna, Tenerife, Spain
    \and 
    Max-Planck-Institut für Sonnensystemforschung, Justus-von-Liebig-Weg 3, 37077 Göttingen, Germany
}

   \date{Received July 30, 2025; accepted November 30, 2025}

 
  \abstract
   {The fine structures of the solar chromosphere, driven by photospheric motions, play a crucial role in  the dynamics of solar magnetic fields. Many have been already identified such as fibrils, filament feet, and arch filament systems. Still, high resolution observations show a wealth of structures that remain elusive.
}
   {We have observed a puzzling, unprecedented chromospheric fibril singularity in close vicinity of a blow-out solar jet and a flaring loop. We aim to understand the magnetic nature of this singularity and the cause of its activity using coordinated high-resolution multi-wavelengths observations.
  }
   {We aligned datasets from Solar Orbiter, SST, IRIS, and SDO. 
   We re-projected the Solar Orbiter datasets to match the perspective of the Earth-based instruments. We performed potential field extrapolations from Solar Orbiter/PHI data.
We analysed the spatial and temporal evolution of the plasma structures and their link with the surface magnetic field. This leads us to derive a model and scenario for the observed structures which we explain in a general schematic representation.}
   {
  We have discovered a new feature, a singularity in the chromospheric fibril pattern. It is formed in a weak magnetic field corridor between two flux concentrations of equal sign, at the base of a vertically inverted-Y shape field line pattern. In this specific case some activity develops along the structure. Firstly a flaring loop at one end, secondly a blow-out jet at the other end, where a coronal null-point was present and associated with a chromospheric saddle point being located onto the fibril singularity. The observations suggest that both active phenomena were initiated by converging photospheric moat flows that exerted pressure on this fibril singularity.}
{}
   \maketitle
%

\section{Introduction}
Fine structures in the solar chromosphere are believed to be 
aligned with 
the magnetic field. Consequently, high-resolution chromospheric observations provide a powerful indirect diagnostic for exploring the geometry and topology of magnetic fields, in particular $\Halpha$ observations serve as an effective proxy for chromospheric magnetograms \citep{Filippov1995}. Recent advances in ground and space-based observational capabilities have markedly improved spatial, temporal, and spectral resolution, offering unprecedented insights into the dynamic fine-scale structure of the solar chromosphere and its coupling to the higher layers of the solar atmosphere.

 Chromospheric fibrils, prominently observed in $\Halpha$, form a dense canopy at chromospheric heights. 
Along these fibrils solar  jets are often observed. These jets are impulsive and collimated plasma ejections \citep{Shibata1992, Nistico2009, Sterling2015, Joshi2024} and often involve plasma at 
cool chromospheric temperatures, commonly referred to as surges \citep{Schmieder1995, Shen2012, Uddin2012, Nobrega2016, Nobrega2017, Shen2017, Joshi2020, Joshi2024b}. 
Magnetic reconnection in the solar corona generates high-velocity hot jets 
visible in X-ray or EUV wavelengths, while reconnection occurring lower in the chromosphere results in cooler \Halpha\ surges. These processes and different types of jets are described in detail in the reviews of, e.g., \citet{Raouafi2016}, \citet{Shen2021}, and \citet{Schmieder_Joshi2022}. 
These jets have been broadly classified as standard and blow-out jets \citep{Moore2010, Shen2012, Sterling2015, Chandra2017a, Kumar2019, Schmieder2022}. Blow-out jets are a more dynamic subclass of solar jets, arising from the eruption of small filament eruptions with some events showing associations with coronal mass ejections \citep{Shen2012, Duan2019, Shen2019, Joshi2020ApJ, Duan2024}. These jets typically originate from a distinct magnetic topology known as an ``X-point'' or ``null point'', where magnetic field lines converge or diverge \citep{Pariat2009}. This configuration facilitates the rapid release of magnetic energy, driving the formation of a jet. 
In an insightful study by \citet{Titov2011}, reconnection along a null line was explored.
In this scenario, instead of an open-field corridor, different reconnection regions can be connected by a singular line corresponding to the separatrix footprint. In the presence of magnetic null points, the connectivity of the magnetic field is determined by the skeleton of the field configuration. When nulls vanish below the photosphere these skeletons transform into quasi-separatrix layers \citep{Restante2009}, where current sheets can naturally form \citep{Aulanier2005}.

Recently, the fine structures related to jets have been observed using the advanced instrumentation of Solar Orbiter \citep{2020A&A...642A...1M}. High-resolution observations from the Extreme Ultraviolet Imager \citep[EUI;][]{2020A&A...642A...8R} have enabled the study of these phenomena on much smaller scales \citep{Mandal2022,Panesar2023,Nobrega2025}, resolving features down to spatial scales of a few hundred kilometres \citep{Chitta2023, 2025A&A...696A.125P}. 
Solar Orbiter observations, together with chromospheric data from ground-based instruments, are crucial for examining whether solar jets always follow to the conventional reconnection scenarios. This motivated us to conduct a coordinated observational campaign using multiple instruments, during which we detected repeated solar jets. What is particularly striking is the chromospheric environment, where fine-scale fibrils emerge and play a crucial role in the propagation of brightenings near the base of a blow-out jet. 
We present observations where the fibril pattern shows an unusual “parting hairstyle” appearance, giving it a distinctive and singular character. The morphology and orientation of chromospheric fibrils allow us to infer the magnetic field geometry in a region 
where direct, high-resolution magnetic field measurements are notoriously difficult to obtain.

   \begin{figure}[t!]
   \centering
   \includegraphics[width=0.48\textwidth]{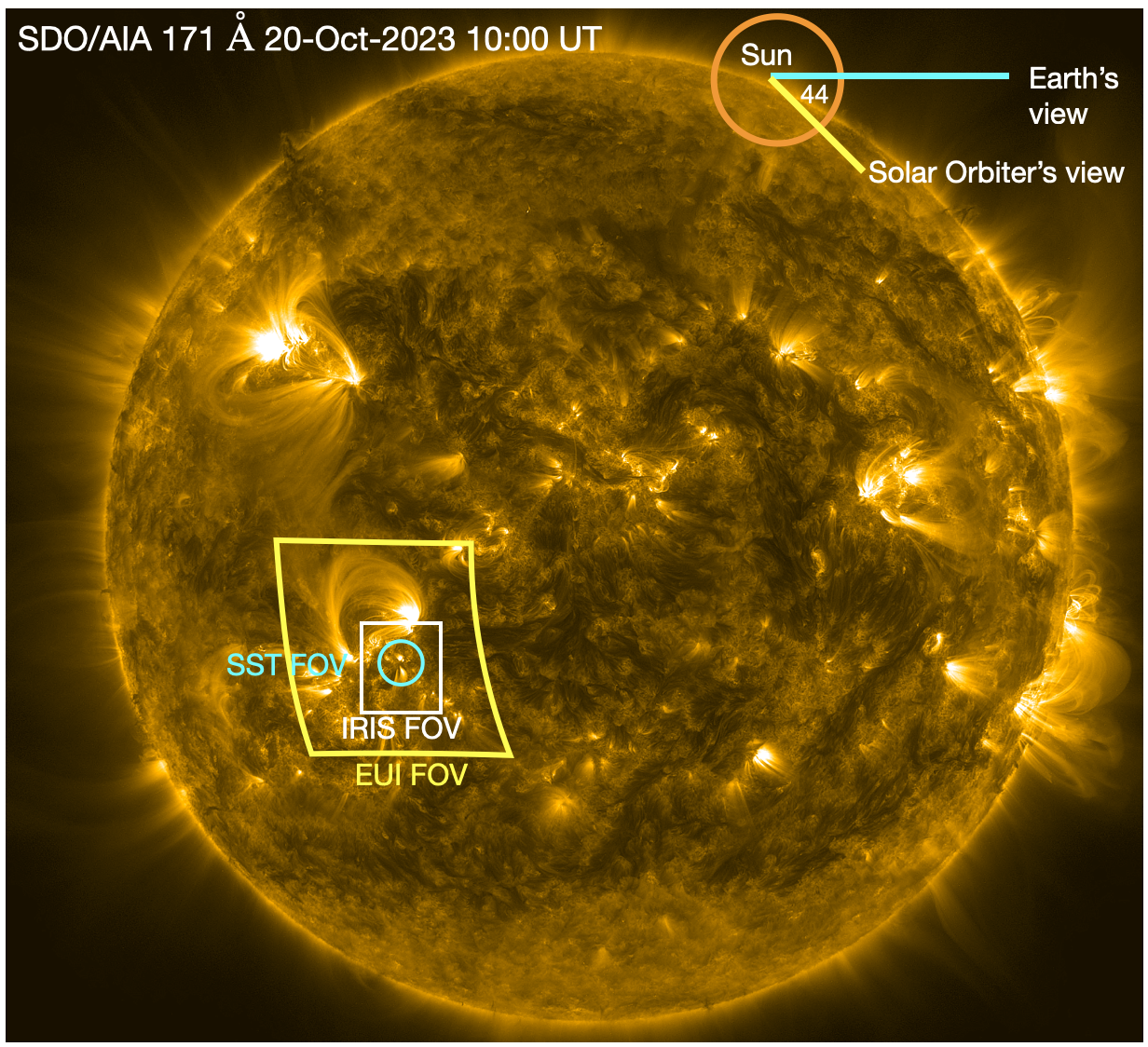}
   \caption{Outline of the areas covered by the different instruments drawn on a full disk AIA 171~\AA\ image. The diagram in the top right indicates the different viewing angle (44\degr) and heliocentric distance (0.41 AU) of Solar Orbiter as compared to SST, IRIS, and SDO which all observe along the Earth-Sun viewing line.
}
              \label{fig:AIA_SolO}%
    \end{figure}
%

  
%
\begin{figure*}
\centering
\includegraphics[width=\textwidth]{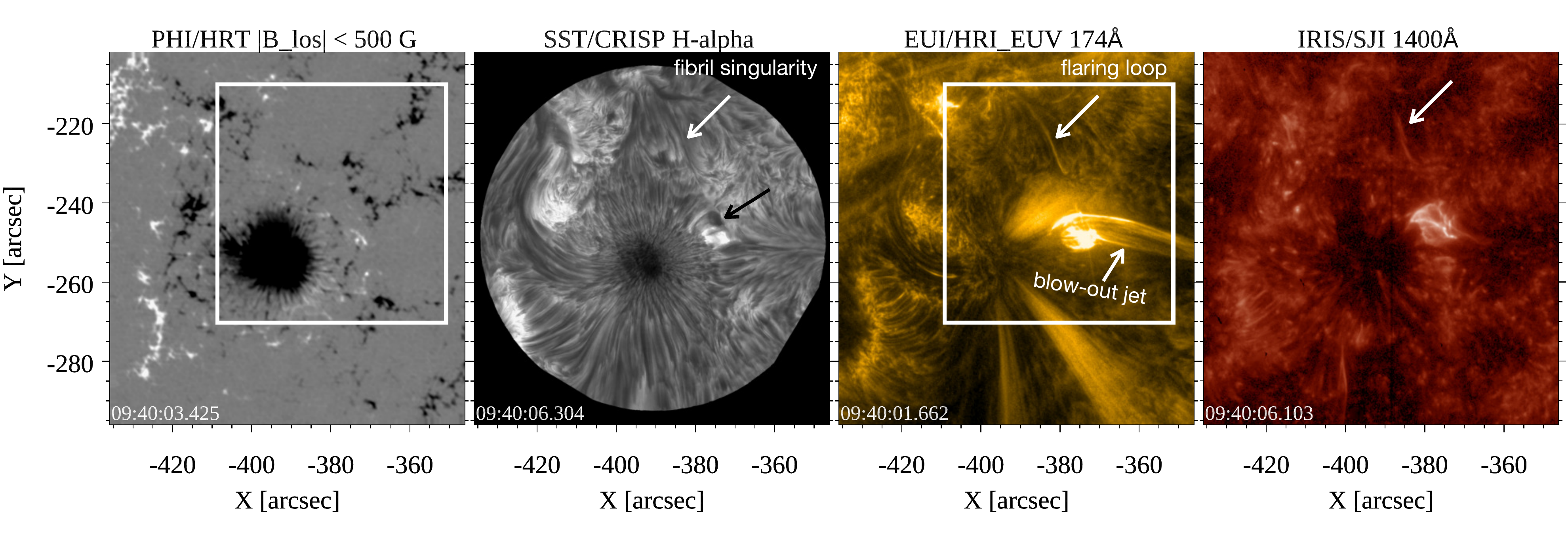}
\caption{Observations from Solar Orbiter (reprojected to Earth's view), SST, and 
IRIS showing a fibril singularity and a blow-out jet around 09:40 UT.
The Solar Orbiter/PHI magnetogram (left panel) shows the sunspot's moat flow, containing small-scale magnetic patches (see the attached animation).
The black arrow in the SST panel points to the chromospheric material ejected from the peeling of the arch filament system. 
The white box shows the FOV of Fig.~\ref{fig:zoom}. 
This figure is associated with an online animation 
(\url{https://drive.google.com/file/d/1O4EYS0pFC0b3BJj3wQm26zneTUkW6_Ev/view?usp=sharing}).}
\label{fig:context}%
\end{figure*}


\begin{figure*}
   \sidecaption
   \includegraphics[width=12cm]{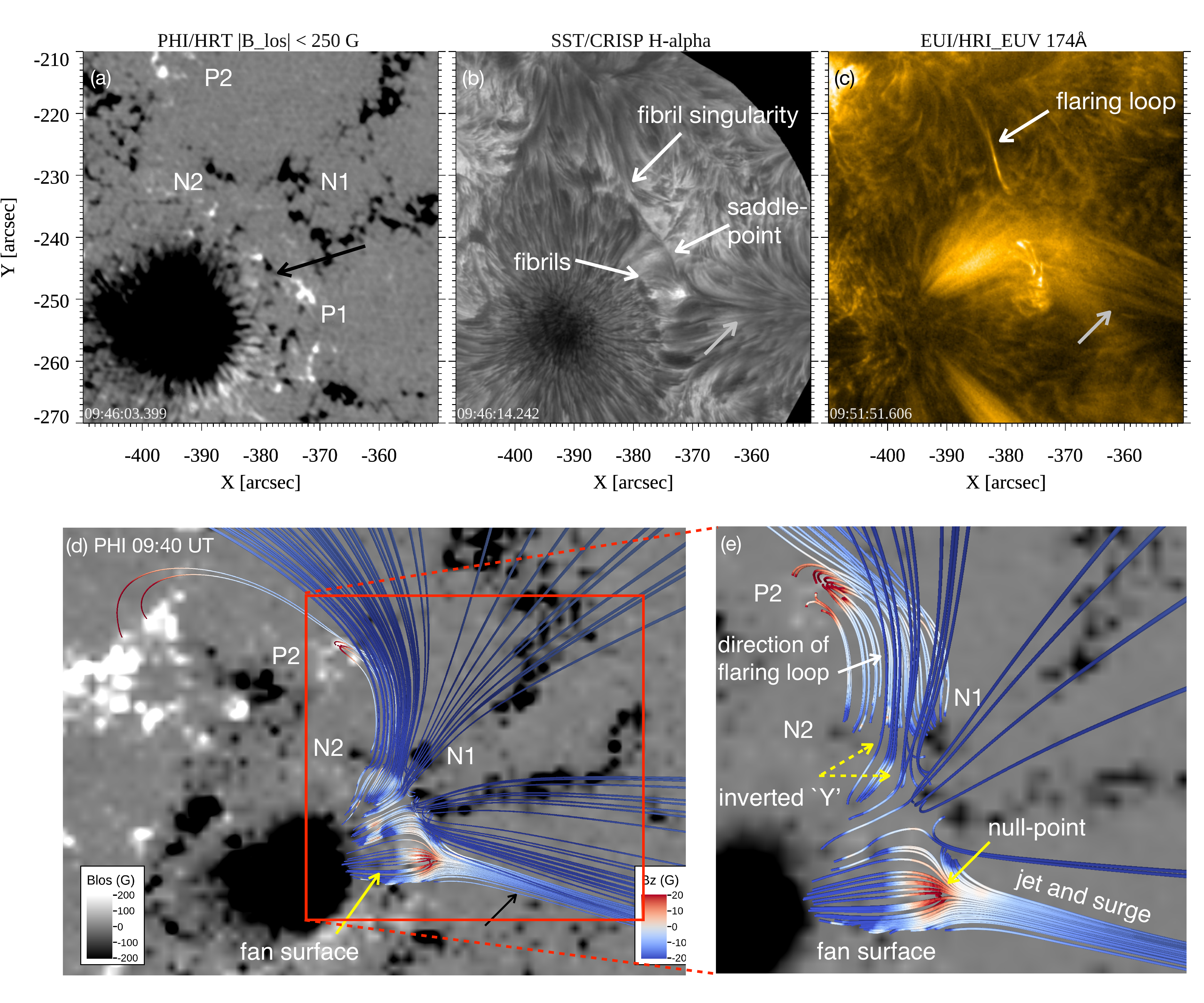}
\caption{
A zoomed-in view on the area outlined by the white box in 
of Fig.~\ref{fig:context}.
A strong moat flow is observed near the jet base area and is marked with the black arrow in panel (a). Four main positive and negative polarity patches are labelled as P1, P2 and N1, N2 respectively.
The blow-out jet in the coronal EUV 174~\AA\ image, associated with a chromospheric surge in \Halpha, is indicated with the gray arrow in panels (b) and (c). Fibrils and the `X'-shaped saddle point structure are marked in the \Halpha\ image. A flaring loop associated with the chromospheric fibril singularity is shown in EUI and \Halpha. Potential field extrapolation on the PHI magnetogram is shown in panel (d-e). In panel (d), the yellow arrow points the jet base which consists of a fan-spine configuration with a null point and associated outer spine towards the west (black arrow). A zoomed-in version is shown in panel (e), where the several inverted `Y' structures have been shown from the weak magnetic corridor between N1 and N2.}

\label{fig:zoom}%
\end{figure*}

   \begin{figure}[t!]
   \centering
   \includegraphics[width=0.49\textwidth]{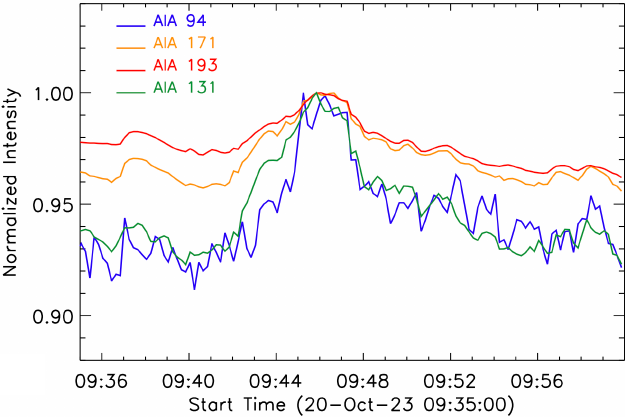}
   \caption{Intensity light curves at the flaring loop location in different AIA channels. 
   The curves display a flare-like evolution, with a rapid heating phase that is most pronounced in the hotter channels and a cooling phase after the peak around 09:46 UT that is slower in the cooler channels.
   The position of the box used to compute the intensity is indicated in Fig.~\ref{fig:aia+eui_jet}.
}
              \label{fig:lightcurve}%
    \end{figure}
%

\section{Observations and methods}
\label{sec:observations}

On 20 October 2023, active region (AR) 13468 was observed by a unique combination of observatories: Solar Orbiter at a heliocentric distance of 0.41~AU, the Interface Region Imaging Spectrograph 
\citep[IRIS;][]{Pontieu2014}, 
the Swedish 1-m Solar Telescope 
\citep[SST;][]{Scharmer2003} 
on La Palma, 
and the Solar Dynamics Observatory 
\citep[SDO;][]{Pesnell2012}. 
An extensive range of the active region atmosphere was covered with high resolution coronal imaging by EUI/High Resolution Imager (HRI) 174~\AA, transition region imaging by the IRIS slit-jaw channel SJI 1400~\AA, and chromospheric \Halpha\ and photospheric wideband imaging by the SST.
Solar Orbiter's Polarimetric and Helioseismic Imager \citep[PHI;][]{2020A&A...642A..11S} 
provided photospheric magnetic field maps 
(see Fig.~\ref{fig:AIA_SolO} for wider context and Fig.~\ref{fig:context}). 
Details on the observations and instruments are provided in Appendix~\ref{app:observations}.
In order to investigate the magnetic field topology 
we performed a potential field extrapolation on the photospheric magnetic field data from PHI.
%

We performed a Fast Fourier Transformation based potential field extrapolation \citep{1972SoPh...25..127N, 1981A&A...100..197A} on the photospheric $B_\mathrm{LOS}$ data from PHI. This choice was made to circumvent challenges and inaccuracies arising from the ambiguity resolution of the full vector magnetic field, allowing us to work with a straightforward and well-established potential field model.
The extrapolation is performed over the full PHI FOV resulting in a box size of 1536 $\times$ 1536 $\times$ 384 grid points, which approximately corresponds to a domain size of 
228~Mm $\times$ 228~Mm $\times$ 57~Mm.
The bottom boundary for extrapolation was also checked to ensure flux balance, allowing the resultant extrapolated magnetic field to closely satisfy the divergence-free condition. The mean value of $B_\mathrm{LOS}$ for this region is $-2.3$~G, which is well below the noise level of about 10~G. 
For visualization of the extrapolated magnetic field lines in 3D, we used the VAPOR software \citep{2019Atmos..10..488L}.
In Fig.~\ref{fig:zoom}, we traced the magnetic field lines near the base of the jet through bidirectional field line integration by randomly placing the seed points in a small volume with a bias towards smaller values of $|B|$. 
This allowed us to trace the field lines closely following the separator connecting the null points.

\section{Chromospheric singularity, flaring loop and blow-out jet}
In this section, we describe two distinct events that occurred in AR 13468: a blow-out coronal jet, and a distinct flaring loop along a long filamentary structure that we refer to as an active fibril singularity, whose characteristics and origin are explored in detail below.

The blow-out jet is found in close proximity to the negative polarity sunspot: a region that produced multiple jets over time. In particular, we focus on the blow-out coronal jet launched around 09:37 UT, as it was simultaneously observed by the SST. The PHI magnetograms clearly show the moat flow surrounding the sunspot; the persistent outflow of magnetic patches away from the sunspot (see Fig.~\ref{fig:context} and associated animation).  
The moat flow transported magnetic flux towards a strong positive polarity patch near $(x,y)=(-370\arcsec,-250\arcsec)$. 
Figure~\ref{fig:moatflow} illustrates how a negative polarity patch collides with this positive patch prior to the start of a blow-out jet. The base of the jet consisted of arch filament systems (AFS) that were oriented perpendicular to the small negative polarities of the moat flow and the positive boundary. This AFS is clearly visible in the  \Halpha\ observations (marked with the black arrow in Fig.~\ref{fig:context}). 
After the AFS became visible in \Halpha, the jet base became bright in EUI~174~\AA\ and in IRIS SJI 1400~\AA\ \ion{Si}{iv}. The blow-out jet became very prominent in EUI 174~\AA\ and was also visible across all other temperature diagnostics.
The EUI images show narrow threads and fine structures in the jet. There are also distinct, high-contrast dark structures inside the blow-out jet that can be associated with the surge that is prominently visible in the \Halpha\ image (gray arrow in Fig.~\ref{fig:zoom}). This surge might be produced by a peeling of the AFS \citep{Joshi2024b} or due to a mini-filament eruption process \citep{Sterling2015, Shen2021}.

Very interestingly, we observe the appearance of a very thin structure that extends towards the North, orthogonal to the jet's direction of propagation. We refer to this structure as a flaring loop in the EUI 174~\AA\ panel in Fig.~\ref{fig:zoom} where it is particularly bright. 
To investigate the nature of the flaring loop, we analysed multi-wavelength observations from SDO/AIA in the 171, 193, 131, and 94~\AA\ channels, focusing on the temporal evolution of the intensity at a selected location within the loop. The normalized intensity light curves exhibit an asymmetric flare-like profile, characterized by a distinct heating phase followed by a gradual cooling phase (see Fig.~\ref{fig:lightcurve}). The intensity was measured within the box indicated in Fig.~\ref{fig:aia+eui_jet}, over the period from 09:35 to 10:00 UT. All AIA channels show an intensity peak around 09:46 UT, followed by a rapid decline in intensity after 09:48 UT, consistent with post-flare cooling. 
The emission in the hotter AIA channels (94 \AA\ and 131 \AA) rises earlier and decays more rapidly than that in the cooler channels, consistent with earlier findings about the flare plasma \citep{Sun2013, Dai2013}.
This flaring loop structure is also present in other diagnostics: fainter in transition region SJI 1400~\AA\ \ion{Si}{iv} (see Fig.~\ref{fig:context}) and as a long and narrow dark fibril in the chromospheric \Halpha\ observations. 
Over the 100-minute sequence of the observation (see animation associated with Fig.~\ref{fig:context}), three flaring loops and three jets are observed along the singularity, showing their recurrent behaviour (see Fig.~\ref{fig:recurrence}). We are focussed on the first flaring loop around the same time as of the blow-out jet (09:40 UT).
This flaring loop seems to appear from the main jet base, which is the region where the moat flow collides with the large positive polarity patch in the PHI magnetograms. 
It appears along a weak magnetic corridor between two large negative polarity patches that are marked N1 and N2 in Fig.~\ref{fig:zoom}. Just below the flaring loop and above the jet base, an X-like structure or a ``saddle-point'' is visible in chromospheric observations (indicated by a white arrow in Fig.~\ref{fig:zoom}b). The term saddle-point is inspired by \citet{Filippov1995}.

Examining the evolution of the magnetic field in the PHI magnetograms, we observe several magnetic polarities moving from the large sunspot toward the prominent negative polarity patch N1. These magnetic polarities labeled as C1, C2, C3, and C4 in Fig.~\ref{fig:phi_evolution}, illustrate the dynamic evolution of small-scale polarities near the main sunspot. The small polarities appear to emerge and propagate through a weak magnetic corridor (all along between N1 and N2), leading to enhanced flux convergence. For instance, the displacement of polarity C1 between 09:05 UT (Fig.~\ref{fig:phi_evolution}a) and 09:55 UT (Fig.~\ref{fig:phi_evolution}f) shows a clear northward motion, as the arrow tip approaches the negative patch at the later time. Similarly, the motion of c4 is evident between panels (e) and (f), while c2 exhibits the fading of a small positive patch at 09:15 UT, which reappears about 40 minutes later as a more circular feature that has migrated toward the weak magnetic corridor. Consequently, magnetic flux converges more intensely in the singularity’s vicinity.
\section{Key findings from observations and magnetic modeling}
A key finding of this study is the identification of the active chromospheric singularity, along which a thin flaring loop oriented perpendicular to the blow-out jet (see the animation at 09:36:02). 
The flaring loop observed around 09:40:01 is both spatially and temporally aligned with the fibril singularity detected in $\Halpha$ observations. A similar alignment is again evident later at 09:53:41 and 10:45:33 with no other activity occurring in the surrounding region. Therefore, we infer that this phenomenon is closely linked to the fibril singularity, which is located within the weak magnetic field corridor between N1 and N2.
Morphologically, the fibrils seem to converge almost horizontally on both sides of this singularity, while along the line itself no fibrils are visible, at least not from above. This makes the line stand out as a unique feature, seemingly devoid of fibrils.

Potential field extrapolation confirms that the blow-out jet originates from a configuration with a characteristic fan-spine topology (with a null-point location), with open field lines extending westward. 
The magnetic field rooted in the main polarities N1 and N2 is vertical with a weak magnetic field corridor between them. These small positive patches are observed in the PHI magnetograms; however, these small magnetic patches are too weak to result in extrapolations that give unambiguous evidence that there exists a null-like location. Notably, multiple inverted `Y' shaped structures appear within this weak magnetic corridor (see Fig.~\ref{fig:zoom}d,e) through the long open field lines  closing at the positive polarity. The long dark filamentary structure in the SST observations represents a chromospheric fibril singularity. A chromospheric saddle-point at one end of the fibril singularity is visible in SST \Halpha\ images in Fig.~\ref{fig:zoom}b.
Given the confirmation from observations and extrapolation, we suggest that the brightening propagation towards North along the flaring loop follows this path, driven by converging motions that exert pressure on this region. 

\begin{figure}[t!]
\centering
\includegraphics[width=0.5\textwidth]{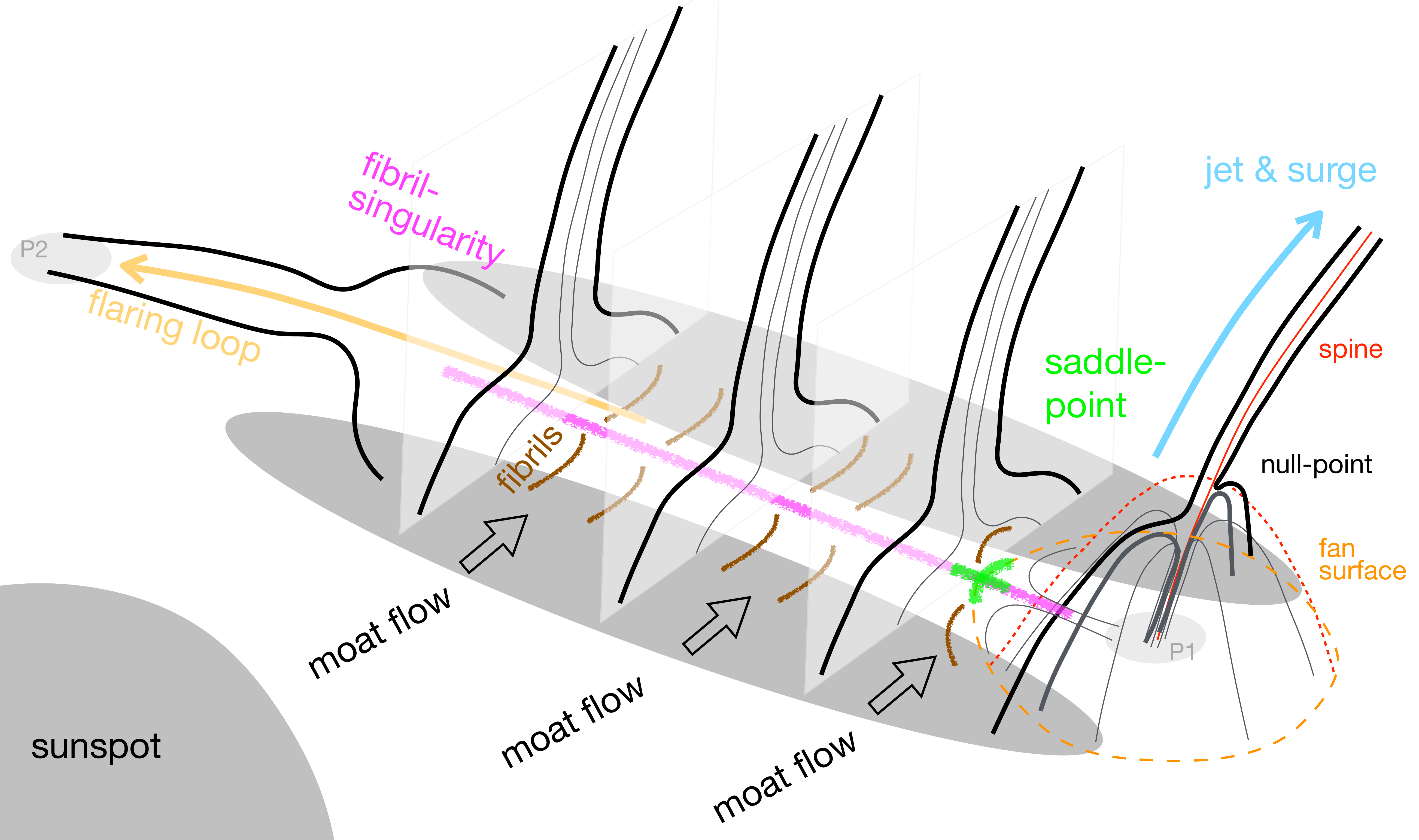}
\caption{
Schematic representation of the observations. The two colliding regions of negative polarity are overlaid with chromospheric fibrils, represented in brown. The flaring loop along the fibril singularity is marked by a curved orange arrow going to the left. The fibril singularity line is shown in magenta with a saddle-point at one end shown in green. A saddle-point is shown  at the boundary of fan surfaces and the fibril singularity. Multiple inverted `Y' structures along the fibril singularity are illustrated with parallel black colour behind the main null-point structure at the jet base.}
\label{fig:sketch}%
\end{figure}
  
Figure~\ref{fig:sketch} presents a schematic illustration of the magnetic environment of this case. However, this sketch is not aimed at being limited to explain this specific observation. The moat flow from the sunspot toward the surrounding negative polarity regions is indicated by black arrows. Two dominant magnetic polarities (gray ellipses) are connected via chromospheric fibrils. 
Some of these horizontal magnetic fields correspond to some fibrils observed in \Halpha\ observations along the weak field corridor.  However, we do not see such fibrils between the two-flux concentration N1 and N2 in the North. It is arguable that, given the geometry of the photospheric Bz, this horizontal field distribution should be present all over the magnetic corridor, which we simplified in the sketch in Fig.~\ref{fig:sketch}.
Orange dashed lines depict the fan surface of the null-point associated with the blow-out jet. Consistent with the magnetic extrapolation, this surface is anchored at the jet base over the positive polarity P1. 
A series of parallel inverted `Y' shaped structures appear rooted in a similar configuration to the main null, forming a complex magnetic topology. The fibril singularity, indicated by the magenta line, runs through this weak magnetic corridor between N1 and N2, intersecting the chain of inverted `Y' shaped structures. 
This fibril singularity is a path along which the horizontal magnetic field emanating from N1 and N2 converge and meet. Hence it is a unique and discontinuous linear structure in chromospheric fibrils.
The brightening propagates along the flaring loop and appears to be connected to a small positive polarity at the top, as identified in the PHI magnetogram (Fig.~\ref{fig:zoom}a,d, and e). This connection is also supported by the extrapolated long closed loops.

In summary, we propose the following scenario: continuous moat flows from the large negative sunspot result in converging magnetic flows between polarities of equal signs N1 and N2. This leads to the compression of the weak magnetic field between  N1 and N2. It is arguable that this flow should compress the chromospheric fibril singularity, and thus should result in the formation of a current layer along the inverted `Y' shaped structures. The continuous moat flow also formed a  null-point structure with a blow-out jet ejection. The null-point and the active fibril singularity are magnetically separated from the saddle-point  observed in the chromosphere.

 

\section{Conclusion}
We have discovered a new feature, an active singularity in the chromospheric fibril pattern, formed in a weak magnetic corridor. An observed saddle-point at the chromospheric heights formed at one end of this singularity and  separates the field lines that tend to converge toward the null-point at the base of a blow-out jet from those that are directed toward the fibril singularity.  Magnetic reconnection is triggering at the interface between the fan separatrix and this  saddle-point. This reconnection gives the separator field line which has been observed as fibril singularity. Our observations support the idea proposed by \citet{Filippov1995} that topological singularities in chromospheric structures are closely associated with solar activity. The `X'-type structure (saddle-point) observed at the chromospheric level has been reported in some ground-based observations and interpreted as the projection of a coronal null point \citep{Shen2019}. Indeed, magnetic field lines on the edge of the fan footpoint can display a saddle-point structure especially when the field lines outside the fan has inverted `Y'-shape structures.
In this configuration, where a minimal magnetic flux is present between the two concentrations, the configuration corresponds to a true separatrix or like  a quasi-separatrix layer.

In our case, the weak vertical magnetic field between the main negative polarities is filled with horizontal fields that converge toward one another at the centre of the weak magnetic corridor. The field lines from the equal sign magnetic polarities diverge locally from their centres, producing horizontal fields along their flanks and forming vertically inverted `Y'-shaped configurations. Although this is not a true topological structure, it nevertheless represents a distinct geometrical one. Such configurations naturally give rise to quasi-separatrix layer footpoints 
\citep[see, e.g., Fig.~5 of][]{Restante2009}. 
 
Our findings highlight the important role of chromospheric fine structures and offer new insights into the coupling between photospheric motions and coronal activity. In the future, coordinated multi-instrument campaigns will be essential to further investigate this concept of fibril singularity and its role in solar jet dynamics.

\begin{acknowledgements}
The Swedish 1-m Solar Telescope (SST) is operated on the island of La Palma by the Institute for Solar Physics of Stockholm University in the Spanish Observatorio del Roque de los Muchachos of the Instituto de Astrof{\'\i}sica de Canarias.
The SST is co-funded by the Swedish Research Council as a national research infrastructure (registration number 4.3-2021-00169).
IRIS is a NASA small explorer mission developed and operated by LMSAL, with mission operations executed at NASA Ames Research Center and major contributions to downlink communications funded by ESA and the Norwegian Space Agency.
SDO observations are courtesy of NASA/SDO and the AIA science teams.
Solar Orbiter (SolO) is a space mission of international collaboration between ESA and NASA, operated by ESA. 
The EUI instrument was built by CSL, IAS, MPS, MSSL/UCL, PMOD/WRC, ROB, LCF/IO with funding from the Belgian Federal Science Policy Office (BELSPO/PRODEX PEA 4000134088); the Centre National d’Etudes Spatiales (CNES); the UK Space Agency (UKSA); the Bundesministerium für Wirtschaft und Energie (BMWi) through the Deutsches Zentrum für Luft- und Raumfahrt (DLR); and the Swiss Space Office (SSO). The German contribution to SO/PHI is funded by the BMWi through DLR and by MPG central funds. The Spanish contribution is funded by AEI/MCIN/10.13039/501100011033/ and European Union “NextGenerationEU”/PRTR” (RTI2018-096886-C5, PID2021-125325OB-C5, PCI2022-135009-2, PCI2022-135029-2) and ERDF “A way of making Europe”; “Center of Excellence Severo Ochoa” awards to IAA-CSIC (SEV-2017-0709, CEX2021-001131-S); and a Ramón y Cajal fellowship awarded to DOS. The French contribution is funded by CNES.
This research is supported by the Research Council of Norway, project number 325491, 
and through its Centres of Excellence scheme, project number 262622. 
The work of GA was funded by the Appel à Proposition de Recherche of CNES/SHM and by the Action Th\'ematique Soleil-Terre (ATST) of CNRS/INSU PN Astro, also funded by CNES, CEA, and ONERA.

R.J. and D.N.S. gratefully acknowledge the Solar Orbiter/EUI Guest Investigator program, where discussions related to this work took place during their two research stays at the Royal Observatory of Belgium.
This project has received funding from Swedish Research Council (2021-05613), Swedish National Space Agency (2021-00116).
A.P. and D.N.S acknowledge support from the European Research Council through the Synergy Grant number 810218 (``The Whole Sun'', ERC-2018-SyG).
N.P. acknowledges funding from the Research Council of Norway, project no. 324523.
The use of UCAR's VAPOR software (\url{www.vapor.ucar.edu}) is gratefully acknowledged. 
We made much use of NASA's Astrophysics Data System Bibliographic Services.
\end{acknowledgements}

\bibliography{reference, luc}
\bibliographystyle{aa}

\begin{appendix}
\section{Observations}
\label{app:observations}

\begin{figure*}[t!]
   \centering
   \includegraphics[width=\textwidth]{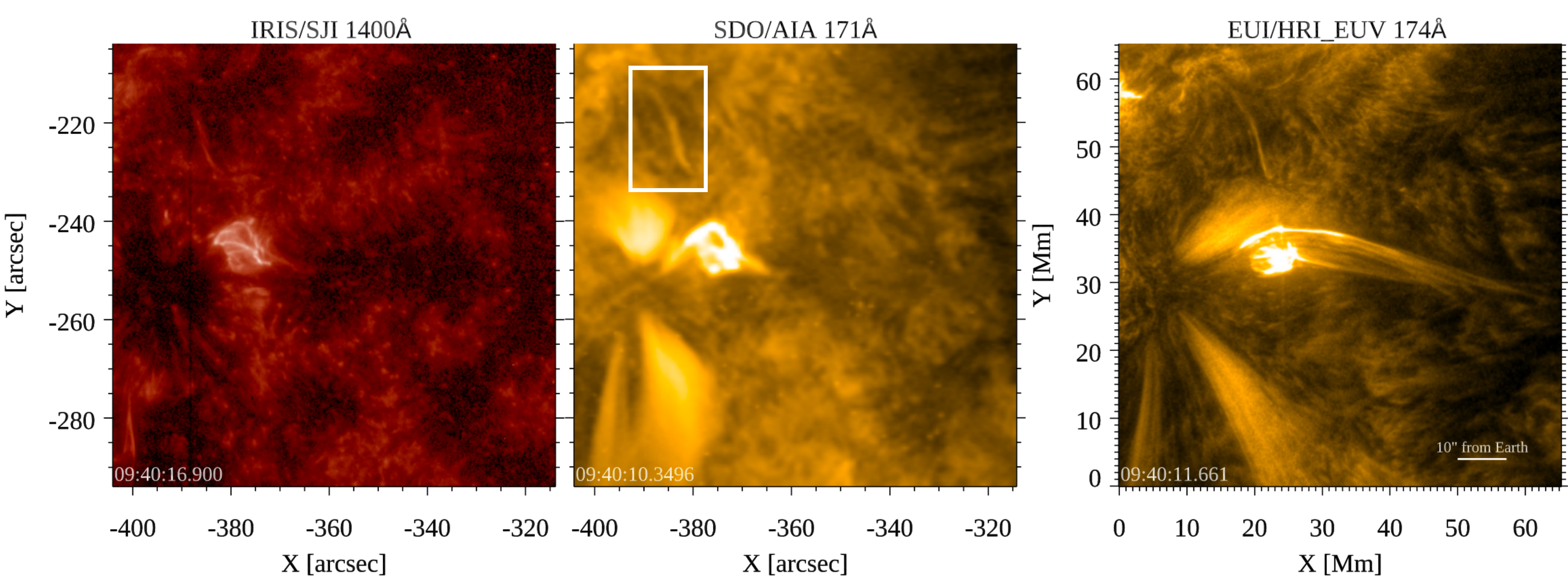}
   \caption{The blow-out jet at the time around maximum extent in IRIS SJI 1400~\AA, AIA 171~\AA, and EUI/HRI$_\mathrm{EUV}$ 174~\AA. The rectangular white box highlighted in the middle panel was used to compute the intensity variation presented in Fig~\ref{fig:lightcurve}.
   The EUI image is shown without reprojection to the Earth viewing direction. 
An animation of this figure is available in the online material
 (\url{https://drive.google.com/file/d/1O4EYS0pFC0b3BJj3wQm26zneTUkW6_Ev/view?usp=sharing}).}
   \label{fig:aia+eui_jet}%
    \end{figure*}

The observations were acquired as part of Solar Orbiter remote sensing window 11 (RSW11) in the period 11 to 22 October 2023. 
On 20 October 2023, Solar Orbiter was running an observing program (SOOP) that had as its main science objective to study flows and waves inside sunspots.
The target was active region NOAA 13468.
Solar Orbiter was at a heliocentric distance of 0.41~au which means that its solar observations have a light travel time difference of about 294~s as compared to Earth. 
As seen from Solar Orbiter, 1\arcsec\ on the Sun corresponded to 296~km (as compared to 723~km as seen from Earth).
Solar Orbiter was positioned at Stonyhurst longitude $-42.2\degr$ and latitude $+6.7\degr$.
The angle between the Earth-Sun and Solar Orbiter-Sun viewing lines was 44\degr. 
Due to this difference in viewing, the apparent size of the classical jet is quite different in the AIA 171~\AA\ and EUI 174~\AA\ images. As seen from Solar Orbiter, this jet reaches a maximum extent of about 50~Mm. In the AIA 171~\AA\ images, the jet is shorter than 18~Mm (see Fig.~\ref{fig:aia+eui_jet}). The long axis of the jet has only a small angle with the line-of-sight from Earth and SDO, while from Solar Orbiter, the jet is viewed much more from the side. For a better comparison, the Solar Orbiter's dataset has been reprojected to the Earth view direction using the reprojection routines of Sunpy \citep{sunpy_community2020}.


%

%

   \begin{figure*}[h!]
   \centering
   \includegraphics[width=\textwidth]{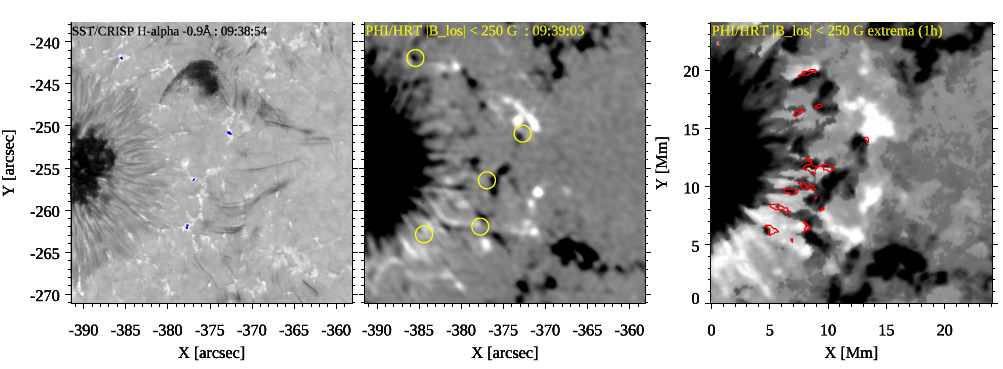}
  \caption{Reconnection regions and rapid motion of photospheric magnetic fields close to the sunspot. In \Halpha\ blue wing (left panel) the reconnection sites are outlined with blue contours, where enhanced \Halpha\ wing emission mark the location of Ellerman Bombs. The intensity at these locations is  1.45 times the intensity of the reference quiet Sun spectrum. The middle panel shows the cotemporal and aligned PHI $B_\mathrm{LOS}$ map. Yellow circles mark the locations of these reconnection points.
  The right panel shows, at each pixel, the extremum of $B_\mathrm{LOS}$ over 1~h before the onset of the jet generated from the original PHI data. Red contours mark pixels that have $|B_\mathrm{LOS}|>100$~G for both polarities during the 1~h sequence. 
  }
   \label{fig:moatflow}%
    \end{figure*}

\begin{figure*}[h!]
   \sidecaption
   \includegraphics[width=12cm]{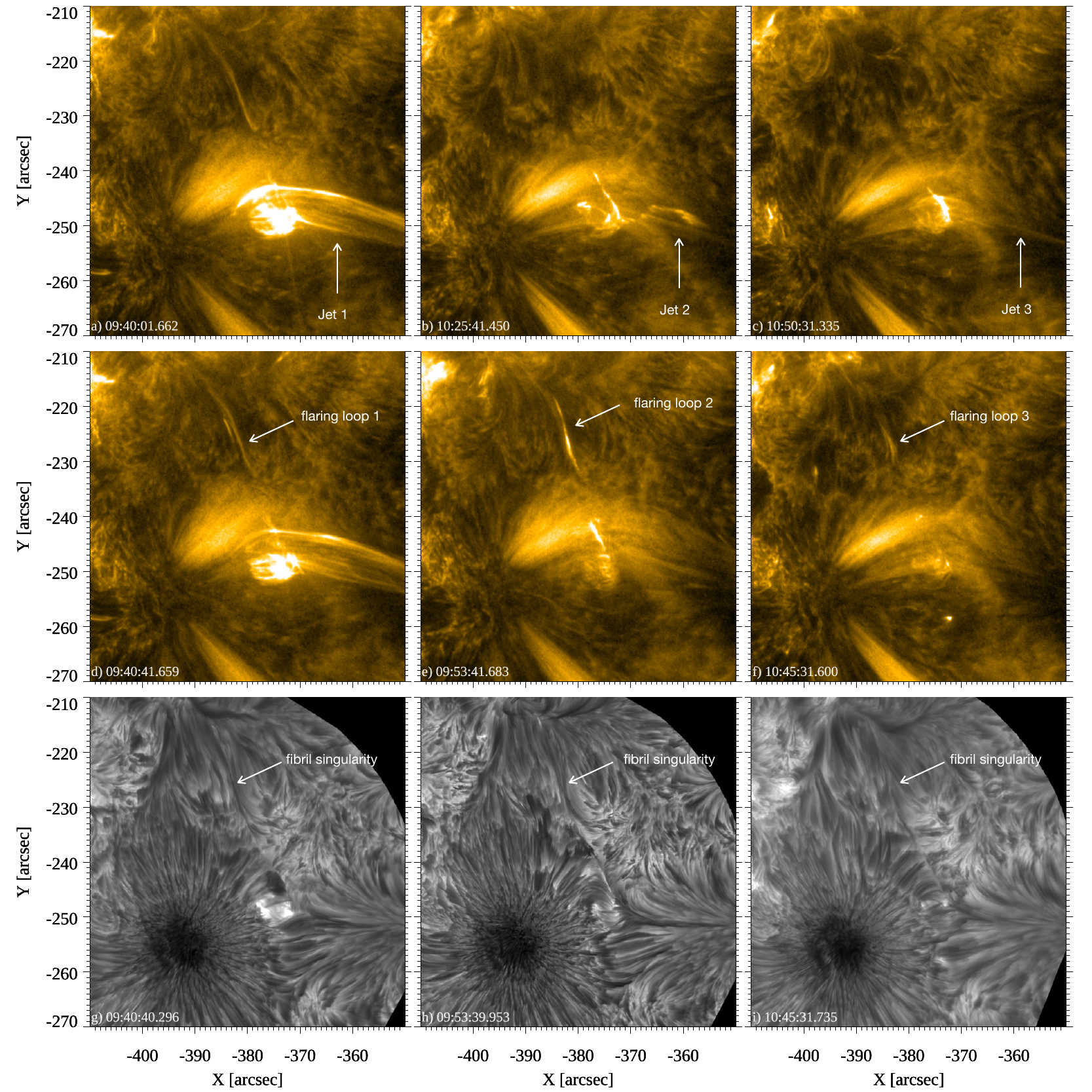}
  \caption{Recurrent blow-out jets and flaring loop along the fibril singularity in EUI/HRI$_\mathrm{EUV}$ 174~\AA\ (top and middle row) and SST/\Halpha\ (bottom row) observations.  These activities along the fibril singularity occur at three different times along the whole duration of the coordinated observations (see the animation associated with Fig.~\ref{fig:context}) as shown in the middle row. Sometimes these are closely associated with the blow-out jets. Our main event is for the Jet 1 and flaring loop 1 around 09:40 UT.
  }
   \label{fig:recurrence}%
    \end{figure*}
   
   \begin{figure*}
   \sidecaption
   \includegraphics[width=11cm]{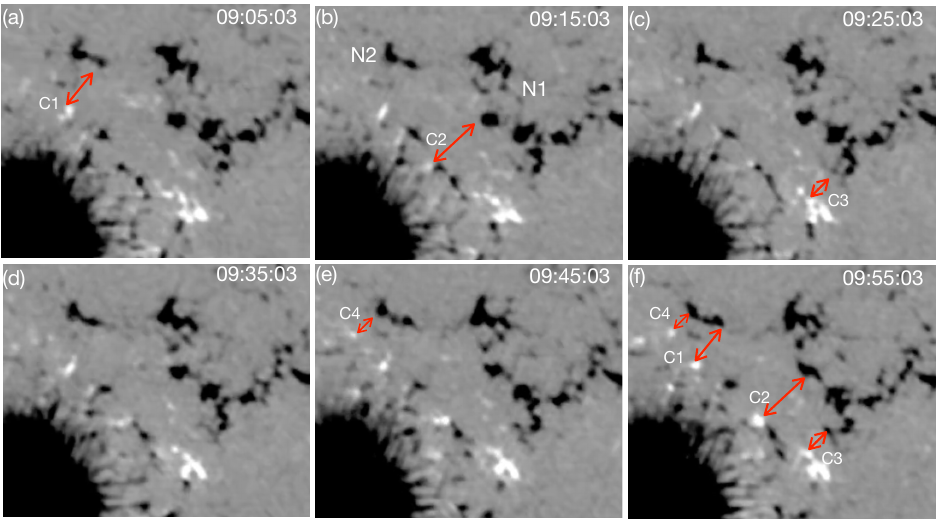}
\caption{Evolution of the magnetic field $B_\mathrm{LOS}$ from PHI observations in the region of interest near the large negative sunspot from 09:05 to 09:55 UT. Several magnetic polarities (labeled C1--C4) originate close to the sunspot. These polarities and the big negative N1-N2 show a motion towards each other over time. 
Four double-headed arrows are drawn in panel (f) such that the arrowheads touch two opposite-polarity patches. The same arrows are drawn on panels (a), (b), (c), and (e) for each of the C1--C4 patches. By comparing these panels with panel (f), it can be seen that the opposite polarities move closer to each other and magnetic flux convergence is increasing.
}
   \label{fig:phi_evolution}%
    \end{figure*}
\subsection{EUI}
\label{app:EUI}

The Extreme Ultraviolet Imager \citep[EUI,][]{2020A&A...642A...8R} 
was running its high-resolution imager at 174~\AA\ (HRI$_\mathrm{EUV}$) at a cadence of 10~s. 
The observations started at 09:00:07~UT and ended at 10:59:37~UT. 
From an Earth-based observer's perspective, the observations started at 09:05:01~UT. 
The HRI$_\mathrm{EUV}$ has a pixel scale of 0\farcs492 which on 20 Oct 2023 corresponded to 146~km in the solar atmosphere. 
The HRI$_\mathrm{EUV}$ FOV covered an area of $297\times297$~Mm (or $411\arcsec\times411\arcsec$ as seen from 1~AU). 
The HRI$_\mathrm{EUV}$ 174~\AA\ channel is dominated by emission from \ion{Fe}{IX} and \ion{Fe}{X} spectral lines. 
The HRI$_\mathrm{EUV}$ 174~\AA\ images that we show here were processed with wavelet-optimized whitening \citep[WoW;][]{2023A&A...670A..66A} 
to enhance small-scale features. 

\subsection{PHI}
\label{app:PHI}

The Polarimetric and Helioseismic Imager \citep[PHI;][]{2020A&A...642A..11S} 
was running its high-resolution telescope \citep[HRT;][]{gandorfer2018high} 
at a cadence of 1~min. 
The observations started at 05:30:03~UT and ended at 11:29:09~UT. 
The HRT has a pixel scale of 0\farcs5 which on 20 Oct 2023 corresponded to 148~km in the solar atmosphere.
The HRT FOV covered an area of $228\times228$~Mm (or $316\arcsec\times316\arcsec$ as seen from 1~AU). 
The PHI data products are derived from Milne-Eddington inversions of spectro-polarimetric observations in the \ion{Fe}{i}~6173~\AA\ line. 
We estimate that the noise level in the level 2 $B_\mathrm{LOS}$ line-of-sight magnetic field maps is about 10~G. This noise estimate was derived from the standard deviation in quiet regions in the data.

\subsection{SST}
\label{app:SST}

At the Swedish 1-m Solar Telescope 
\citep[SST,][]{Scharmer2003}, 
we acquired imaging spectroscopy data in the \Halpha\ line with the tunable filtergraph CRISP
\citep{Scharmer2008}. 
The \Halpha\ line was sampled at 19 line positions with regular steps of 0.1~\AA\ between $\pm$0.9~\AA\ offset from nominal line center. 
The time to complete one spectral scan was 8.9~s. 
Because the seeing was variable and at times of not acceptable quality, the data was recorded in trigger mode: the acquisition of a new spectral line scan was only triggered if the seeing quality was higher than a specific threshold. 
The seeing quality at the SST is measured by the adaptive optics system
\citep{2024A&A...685A..32S} 
in terms of the Fried parameter $r_0$
\citep[also see][]{2019A&A...626A..55S}. 
The trigger threshold was put to $r_0 \ge 6$~cm for the ground-layer seeing. 
The resulting 106~min time sequence has 511 time steps with variable interval. The average cadence is 12.5~s. There are three time gaps that are longer than 1~min (the longest 105~s). 
The Fried parameter varied between 3 and 15~cm (3--26~cm for the ground-layer seeing).
The observations started at 09:05:03~UT. 
After the upgrade of the cameras in late 2022, the CRISP FOV is circular with a diameter of about 87\arcsec\ and the pixel scale is 0\farcs044~pixel$^{-1}$ (32~km). 
The data was processed following the standard SST data reduction pipeline 
\citep{Rodriguez2015, 
2021A&A...653A..68L} 
which includes Multi-Object Multi-Frame Blind Deconvolution 
\citep[MOMFBD,][]{Noort2005} 
image restoration.

\subsection{IRIS}
\label{app:IRIS}

The Interface Region Imaging Spectrograph 
\citep[IRIS,][]{Pontieu2014} 
was running a very-large sparse 2-step raster program (OBSID 3420257611) with an exposure time of 4~s. 
The slit-jaw imager only acquired images in the SJI 1400~\AA\ channel at a cadence of 10~s. 
The pixel scale of the SJI 1400 images is 0\farcs17 (120~km) and the FOV about 166\arcsec\ $\times$ 182\arcsec. 
IRIS observed between 09:06:17 and 10:54:00~UT. 

\subsection{SDO}
\label{app:SDO}

From the Solar Dynamics Observatory 
\citep[SDO,][]{Pesnell2012}, 
we use data from the Atmosphere Imaging Assembly \citep[AIA,][]{Lemen2012} and the Helioseismic Magnetic Imager \citep[HMI,][]{Scherrer2012} instruments for context and comparison. 
AIA observes the full solar disk at a cadence of 12~s in the EUV channels and at a cadence of 24~s in the 1600~\AA\ and 1700~\AA\ UV channels.
The pixel size of AIA is 0\farcs6 (434~km). 
HMI provides maps of the full disk photospheric magnetic field  with a cadence of 45~s and a pixel size of 0\farcs5 (361~km).
We aligned the SDO data to the SST observations using the methods in Solar Soft SSWIDL developed by Prof. R. Rutten\footnote{\url{https://robrutten.nl/rridl/00-README/sdo-manual.html}}.

\end{appendix}

\end{document}